\documentclass[doublecol]{epl2} 

\usepackage{hyperref}

\title{Information content: assessing meso-scale structures in complex networks}
\shorttitle{Information content: assessing meso-scale structures in complex networks} 

\author{M. Zanin\inst{1,2} \and P. A. Sousa\inst{1} \and E. Menasalvas\inst{3}}
\shortauthor{M. Zanin \etal}

\institute{                    
  \inst{1} Faculdade de Ci\^encias e Tecnologia, Departamento de Engenharia Electrot\'ecnica,
Universidade Nova de Lisboa, Portugal.\\
  \inst{2} Innaxis Foundation \& Research Institute,
Jos\'e Ortega y Gasset 20, 28006, Madrid, Spain. \\
  \inst{3} Center for Biomedical Technology, Technical University of Madrid, Campus Montegancedo, 28223 Pozuelo de Alarc\'on, Madrid, Spain\\
}


\abstract{
We propose a novel measure to assess the presence of meso-scale structures in complex networks. This measure is based on the identification of regular patterns in the adjacency matrix of the network, and on the calculation of the quantity of information lost when pairs of nodes are iteratively merged. We show how this measure is able to quantify several meso-scale structures, like the presence of modularity, bipartite and core-periphery configurations, or motifs. Results corresponding to a large set of real networks are used to validate its ability to detect non-trivial topological patterns. }

\begin{document}

\maketitle

\section{Introduction}
In the last decade, complex network theory \cite{albert2002A,albert2002B} has unveiled several topological characteristics that are obiquitous among many real-world systems.
Initially the attention was directed towards two global, {\it macro-scale} network structures, {\it i.e.} {\it small-world} and {\it scale-free} topologies. But soon it was found that complex networks typically possess non-trivial patterns of connectivity at a {\it meso-scale} level, {\it i.e.} between micro and macroscopical scales \cite{almendral2011}, which have been shown to have an important impact on, for instance, spreading \cite{wu2008A,wu2008B} and synchronization \cite{arenas2006A,arenas2006B} processes.

Among the different types of meso-scale structures that have been described, one has received most of the attention: {\it communities}, that is, the organization of nodes in clusters, with many links connecting nodes belonging to the same cluster and comparatively few joining nodes of different clusters \cite{fortunato2010A,fortunato2010B,fortunato2010C}. 
The pervasiveness of a community structure can, in principle, be characterized by quantifying the network {\it modularity} \cite{newman2006}. Yet, this metric suffers from two main drawbacks: first of all, it is {\it a posteriori} metric, in that it can only be calculated after a community structure has been defined. Furthermore, modularity is not robust to the presence of different topological scales, {\it e.g.} when one community is much smaller than the others \cite{danon2006A,danon2006B}. While the concept of modularity can be generalized to include other meso-scale structures, as for instance bipartite networks \cite{guimera2007}, it still inherits the previously discussed drawbacks.
Other types of meso-scale structures, important to understand the structure and dynamics of real networks, include {\it motifs}, {\it i.e.} sub-graphs that recur within a network with a frequency higher than expected in random ensembles \cite{milo2002}, and {\it core-periphery}, composed of a densely connected inner core and a set of peripherical nodes sparsely connected with the core \cite{holme2005}.

In this Letter, we address the following question: is it possible to define a single metric able to detect the presence of different kinds of meso-scale structures? We propose a novel metric, called {\it Information Content}, which is simultaneously {\it (i)} capable of detecting generic {\it regularities} in the adjacency matrix of a network, {\it (ii)} {\it a priori} metric, {\it i.e.} not requiring any previous computation like community detection, and {\it (iii)} robust to different topological scales.

The guiding hypothesis here is that important meso-scale structures are associated with regularities in the corresponding adjacency matrix. For instance, in the simplest case of a network with a perfect modular structure, nodes connect to all peers belonging to the same community: the resulting adjacency matrix is composed of four blocks, two containing only ones, two only zeros (see Eq. \ref{eq:comm} below). In this case, erasing nodes within one community causes no loss of information, as their connections are equivalent; thus, measuring the information lost when pairs of nodes are merged can be used as a way of detecting such kind of regularities - and hence meso-scale structures.

Given an initial network, the proposed algorithm identifies the pair of nodes whose merging would suppose the smallest information loss, a quantity which is a function of the number of common links to / from other nodes shared by the pair. 
Once the best pair has been detected, both nodes are merged (thus yielding a network one node smaller), and the quantity of information $I$ lost in the process is calculated. When this process is iteratively repeated, the {\it Information Content} $IC$ of the network is defined as the sum of all $I$s, {\it i.e.} of all information contained in the network. The lower $IC$, the more regular is the link arrangement, indicating the presence of meso-scale structure.

As such, the calculation of the {\it Information Content} can be seen as a type of network renormalization procedure \cite{Radicchi2008A,Radicchi2008B}, characterized by two specific features. First of all, the objective is the estimation of the quantity of information lost in the process, while classical renormalization focuses on how some properties of the system are conserved at different scales. Furthermore, the renormalization transformation is guided by information theory criteria, instead of geometrical (topological) rules.

%

\section{Information Content calculation}
The calculation of the {\it Information Content} starts with a network of $n$
nodes, which is fully defined by its adjacency matrix $\mathcal{A}$, whose elements $a_{ij}$ are equal
to one when a link exists between nodes $i$ and $j$, and zero otherwise. 
The amount of information that would be lost if two nodes were merged together is first estimated for each pair of nodes $k,l$ (with $k \neq l$). This is performed by comparing the connections departing from and arriving at both nodes, {\it i.e.} the vectors $a_{k\cdot}$, $a_{\cdot k}$, $a_{l\cdot}$ and $a_{\cdot l}$, and by creating a new vector $\bm{m}$ of size $2n$, representing the links that should be modified
to recover the connections of node $l$ given the connections of node $k$, and thus the information lost when both nodes are merged together. 
In the first half of $\bm{m}$,
the $i$-th element (with $i \in [1, n]$) is defined as one if $a_{ki} \neq a_{li}$, and zero otherwise, thus accounting for
different outgoing links; the second half of $\bm{m}$ accounts for different incoming
links: thus $m_{i+n}$ (again with $i \in [1, n]$) is set to one when $a_{ik} \neq a_{il}$, and zero otherwise. 
In the two extreme situations, when two nodes either share all links or none, $\bm{m}$ will either take all values $0$ or $1$ respectively.

Once the vector $\bm{m}$ is constructed, the probability of finding an element equal to one (zero) is given
by

\begin{eqnarray}
p_1  = \frac{1}{2n}\sum\limits_{i = 1}^{2n} {m_i }, \\
p_0 = 1 - p_1.
\end{eqnarray}

Finally, the information contained in $\bm{m}$ is assessed through the Shannon's entropy \cite{shannon1949}:

\begin{equation}
I_{kl}  = 2n\left( { - p_0 \log _2 p_0  - p_1 \log _2 p_1 } \right).
\end{equation}

$I_{kl}$ is defined in $[0, 2n]$, being $I_{kl} = 0$ when $p_0 = 1$ or $p_1 = 1$, meaning that all links are respectively equal or different, and $I_{kl}=2n$ when there is no correlation between the links of nodes $k$ and $l$.

\begin{figure}[!tb]

\includegraphics[width=0.2\textwidth]{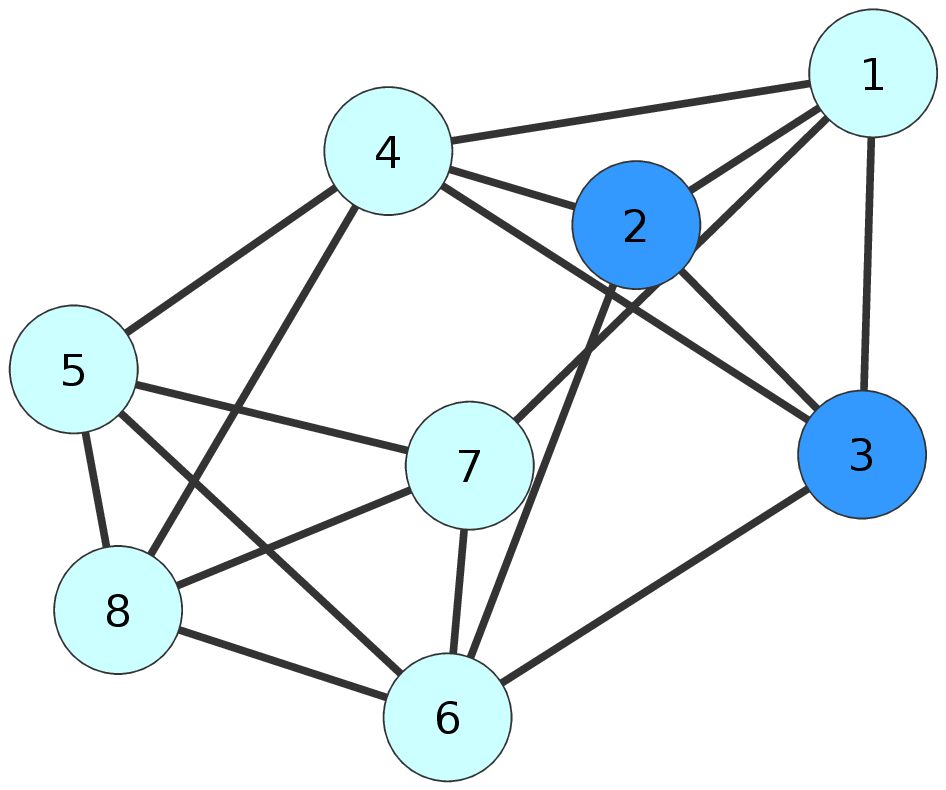}
\hspace{0.03\textwidth}
\includegraphics[width=0.15\textwidth]{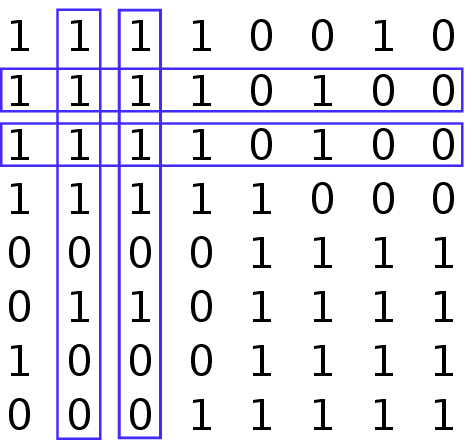}

\includegraphics[width=0.2\textwidth]{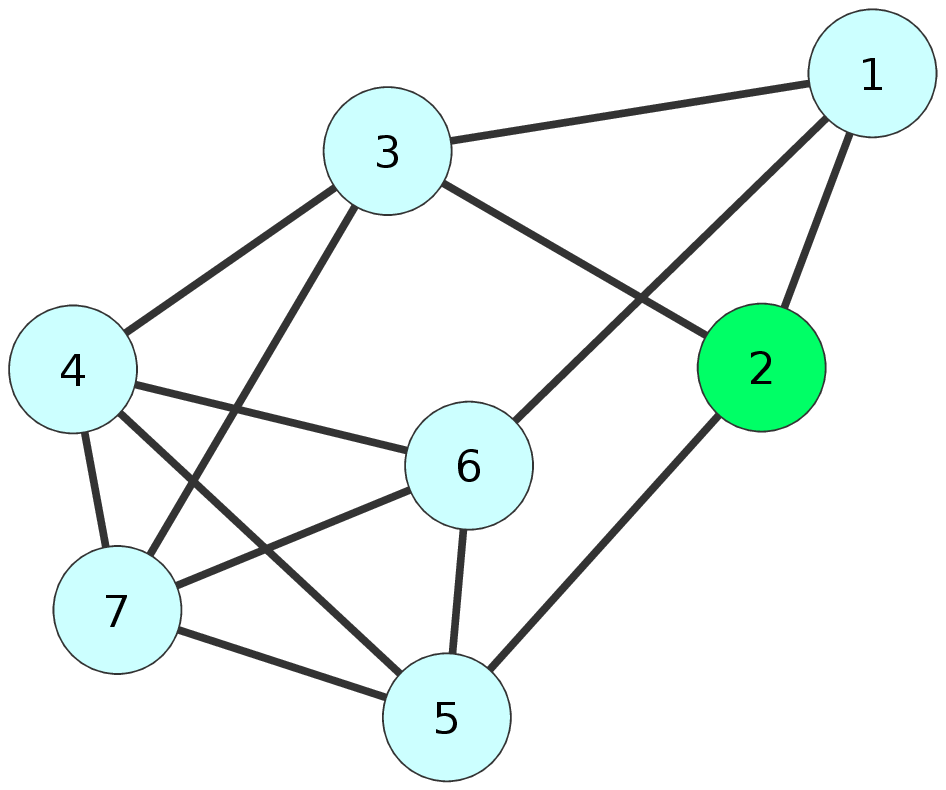}
\hspace{0.03\textwidth}
\includegraphics[width=0.15\textwidth]{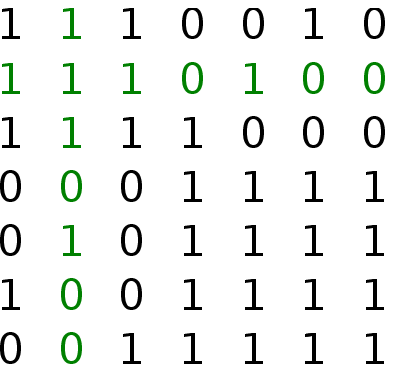}

\caption{(Color online) {\it Example of one iteration of the {\it Information Content} assessment process}.
(Top Left) Initial network, composed of $8$ nodes arranged in two communities (respectively composed of nodes $1-4$ and $5-8$). Notice that nodes $2$ and $3$ (in blue) share the same links. (Top Right) Adjacency matrix of the initial network; blue boxes highlight the four vector of incoming and outgoing links for nodes $2$ and $3$. (Bottom Left) The network after the merging process; the new node 2 (in green) is the result of merging the old nodes $2$ and $3$. (Bottom Right) Adjacency matrix of the resulting network.
\label{fig:1}}
\end{figure}

Once $I$ has been assessed for all possible pairs of nodes, the algorithm identifies the pair
whose merging will suppose minimum information loss. Such pair is then merged by deleting one of its nodes, and the original network is
transformed into a new one composed of $n-1$ nodes (see Fig. \ref{fig:1} for an example). The whole process is then repeated iteratively, until one single node remains.

Each merging step supposes some loss of information (previously denoted by $I_{k,l}$): the {\it Information Content} $IC$ is given by the total amount of information lost as a result of the merging steps leading from the initial network to a single node. Converserly, it can be seen as the amount of information needed to reconstruct the full topology of the network, once it is reduced to a single node, by the merging process.

Two aspects of this metric should be clarified. Firstly, the information included in $IC$ is not complete, as for instance at each step it would be necessary to track which pair of nodes has been merged: yet, the quantity of information required for this is constant, as does not depend on the topology of the network, and is thus discarded. Secondly, the Shannon entropy only provides a lower bound to the quantity of information required to encode vector $\bm{m}$, which may be lower than what required in real applications.

%

\section{The meaning of Information Content}

For a network with a completely random structure, no correlation is expected on average between incoming and outgoing links of any pair of nodes: thus, merging pairs of nodes will result in a nearly maximal $I$, and a maximal $IC$ is expected. This can be used to normalize the {\it Information Content} of any network, such that:

\begin{equation}
IC_{norm}  = IC / \langle IC_{random} \rangle,
\end{equation}

$\langle IC_{random} \rangle$ being the average $IC$ obtained for an ensamble of random networks with the same number of nodes and links of the original graph.

If $\langle IC_{random} \rangle$ provides the upper bound of $IC$, it is easy to find regular structures that will result in a very low {\it Information Content}. Clearly $IC = 0$ both for empty ($a_{ij}=0$, $\forall i, j$) and fully connected networks ($a_{ij}=1$, $\forall i, j$), as merging two nodes would suppose no information loss. More interestingly, the same will occurs with a fully modular network, such that

\begin{equation}
A = \left[ {\begin{array}{*{20}c}
   1 & 1 & {} & 0 & 0  \\
   1 & 1 & {} & 0 & 0  \\
   {} & {} &  \cdots  & {} & {}  \\
   0 & 0 & {} & 1 & 1  \\
   0 & 0 & {} & 1 & 1  \\
\end{array}} \right].
\label{eq:comm}
\end{equation}

\begin{figure}[!tb]
\centering{
\includegraphics[width=0.3\textwidth]{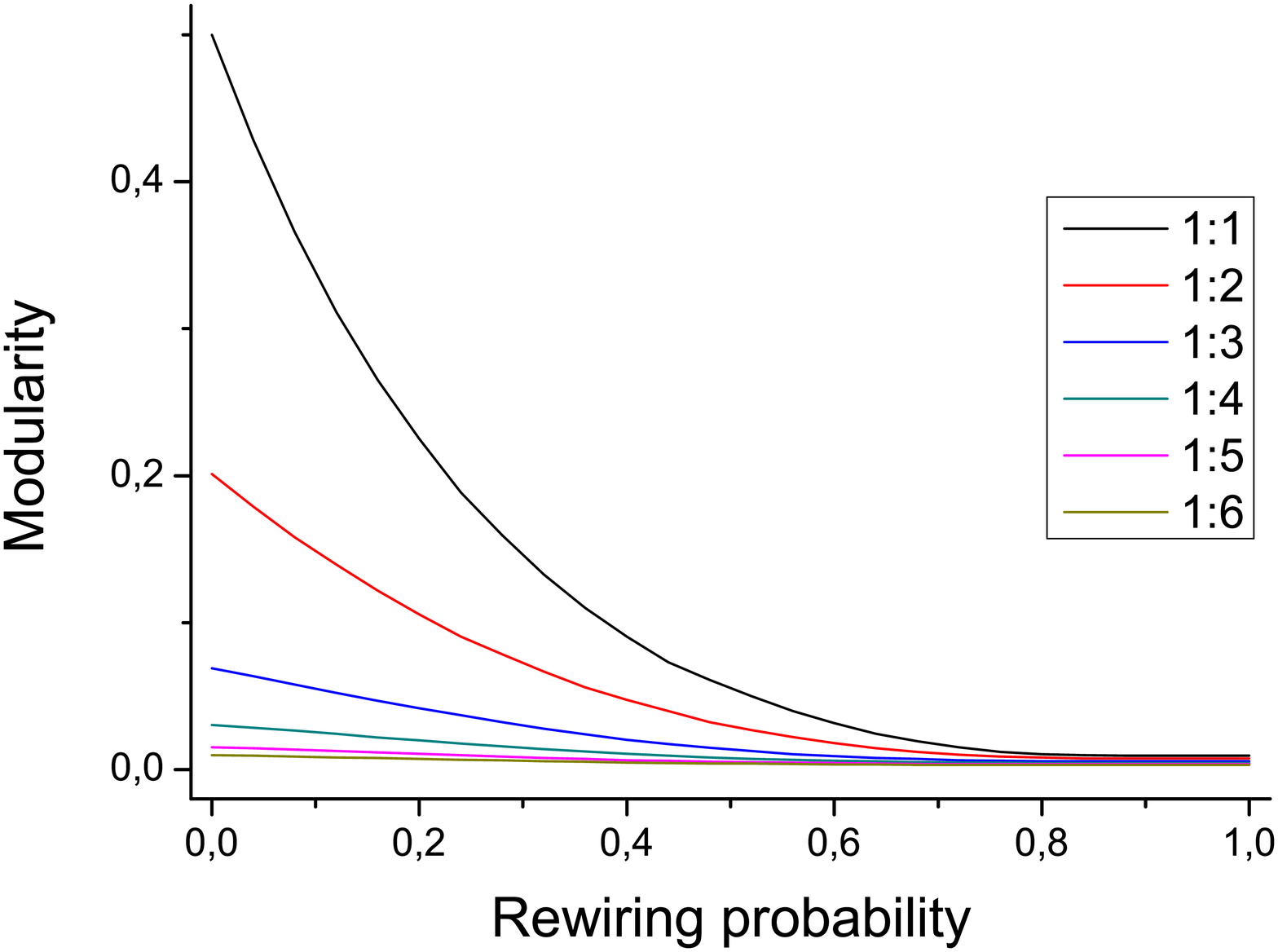}
\includegraphics[width=0.3\textwidth]{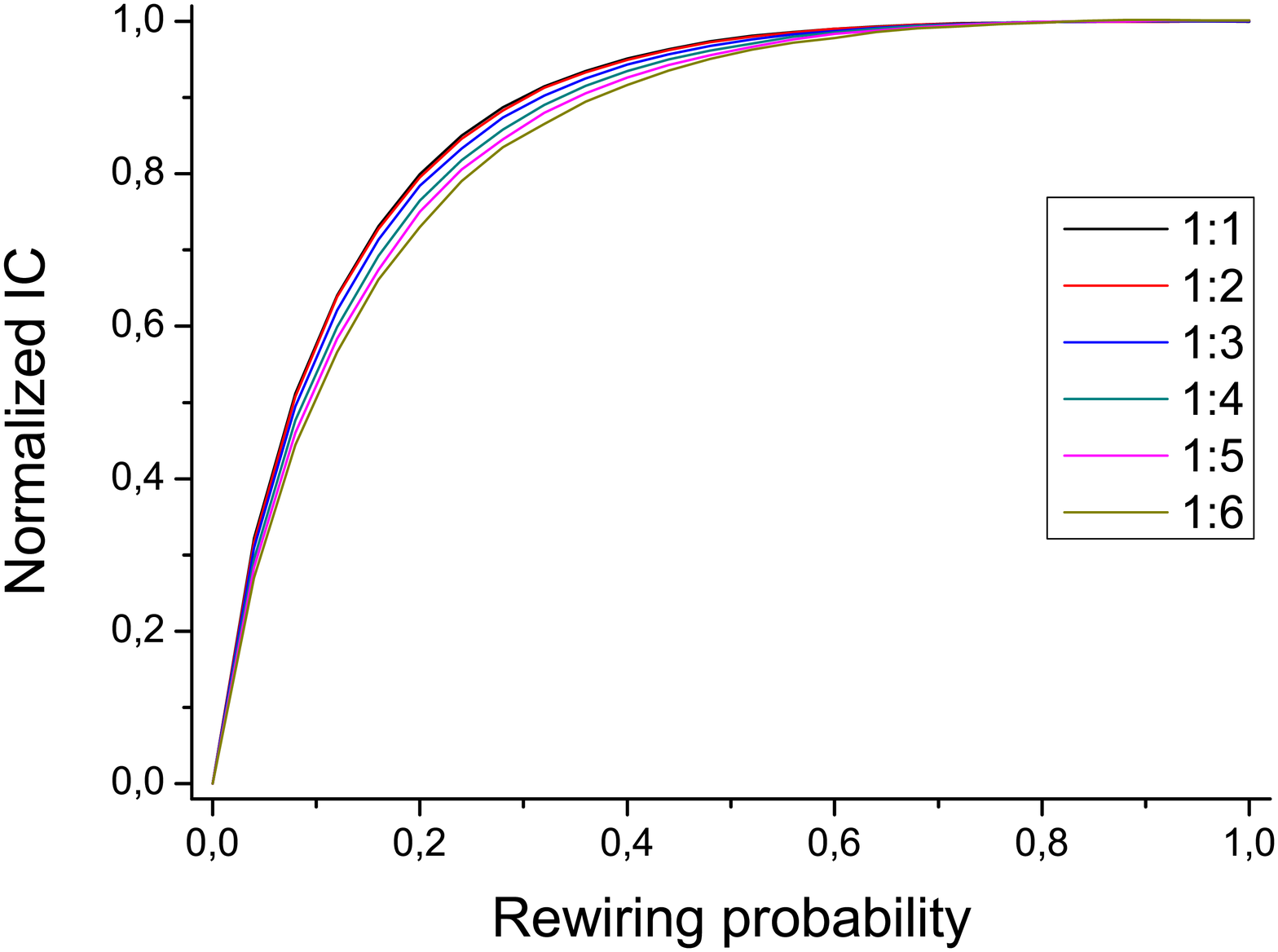}
}
\caption{(Color online) {\it Modularity vs. $IC_{norm}$}.
(Top) Modularity (as calculated with the Blondel's community detection algorithm \cite{blondel2008}) for a network of $400$ nodes organized in two communities. The different lines represent different sizes of the two communities: $1:1$ (black line) two communities of $200$ nodes, $1:2$ (red line) $134$ and $266$ nodes respectively, and so forth. (Bottom) Normalized {\it Information Content} for the same networks. 
\label{fig:2}}
\end{figure}

The fact that all pairs of nodes have either the same or the opposite connections, thus either $p_1 = 0$ or $p_1 = 1$ and $I_{kl} = 0$ for any $k$ and $l$, and $IC = IC_{norm} = 0$, can be used to assess the modularity of a network: moving from a perfectly modular to a random structure, the $IC_{norm}$ smoothly increases from zero to one. 
Contrary to traditional community detection algorithms, $IC_{norm}$ is unaffected by the presence of multiple, widely separated, scales.
Both ideas are demonstrated in Fig. \ref{fig:2}, in which different rewiring probabilities are applied to an initial network of $400$ nodes, comprising two communities of different sizes.

More generally, $IC$ can be used to assess the presence of any regular mesoscale structure. 
Consider for instance a bipartite network, {\it i.e.} networks where nodes belong to two groups, with nodes belonging to one of them being connected only to nodes of the other. The resulting adjacency matrix would thus have the following structure:

\begin{equation}
A = \left[ {\begin{array}{*{20}c}
   0 & 0 & {} & 1 & 1  \\
   0 & 0 & {} & 1 & 1  \\
   {} & {} &  \cdots  & {} & {}  \\
   1 & 1 & {} & 0 & 0  \\
   1 & 1 & {} & 0 & 0  \\
\end{array}} \right].
\end{equation}

Similar results can also be obtained for networks showing a {\it core-periphery} structure, with a densely connected inner core, and a set of peripherical nodes sparsely connected with the core \cite{holme2005}. In this case, merging nodes in the network core will result in low information loss, with a $IC_{norm}$ lower than expected for random graphs.

\begin{figure}[!tb]
\centering{
\includegraphics[width=0.35\textwidth]{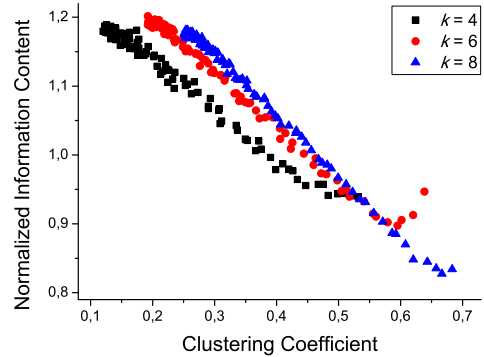}
}
\caption{(Color online) {\it $IC_{norm}$ and Clustering Coefficient}.
Evolution of the $IC_{norm}$ as a function of the Clustering Coefficient. Black squares, red circles and blue triangles respectively correspond to networks with mean degree of $4$, $6$ and $8$.
\label{fig:CC}}
\end{figure}

The previously described meso-scale structures are mainly defined at a global level, in that they affect the overall topology of the network; thus, one may ask if the proposed $IC$ is also effective in detecting more local meso-scales, {\it i.e.} those defined slightly above the single node level. To this aim, we test the measure against networks with high global Clustering Coefficient $CC$, defined as the number of closed triplets (or triangles) over the total number of (both open and closed) triplets. Networks were constructed following the classical method proposed by Watts and Strogatz \cite{watts1998}, {\it i.e.} by starting from regular lattices of fixed degree (thus maximizing the Clustering Coefficient) and by applying a random rewiring process. Results are reported in Fig. \ref{fig:CC}, for networks of 200 nodes and initial degrees of $4$, $6$ and $8$; a clear correlation can be found between $IC_{norm}$ and $CC$, such that the higher the latter, the more regular is the resulting topology, thus yielding low $IC_{norm}$ values.

The Clustering Coefficient can be seen as a special case of {\it motifs}, {\it i.e.} sub-graphs recurring within a network with a higher than expected frequency \cite{milo2002}. Their importance resides in the fact that they can be understood as basic building blocks, each one of them associated with specific functions within the global system \cite{shen2002}. The main difference with complete triangles is that motifs are not necessarily symmetrics nor complete, thus one expects a lower contribution toward creating regular structures in the adjacency matrix. By analyzing the $IC_{norm}$ in random networks as a function of the frequency of appearance of different 3-nodes motifs, a significant correlaction can be found with motifs $3$ ($\rho = -0.7970$, $r^2 = 0.6194$), $5$ ($\rho = -0.7557$, $r^2 = 0.5711$), $7$ ($\rho = -0.7888$, $r^2 = 0.6222$) and $9$($\rho = -0.7415$, $r^2 = 0.5498$) - for the enumeration of 3-nodes motifs, refer to Fig. 1B of Ref. \cite{milo2002}.

%

\section{Application to real networks}

\begin{figure*}[!tb]
\centering{
\includegraphics[width=0.4\textwidth]{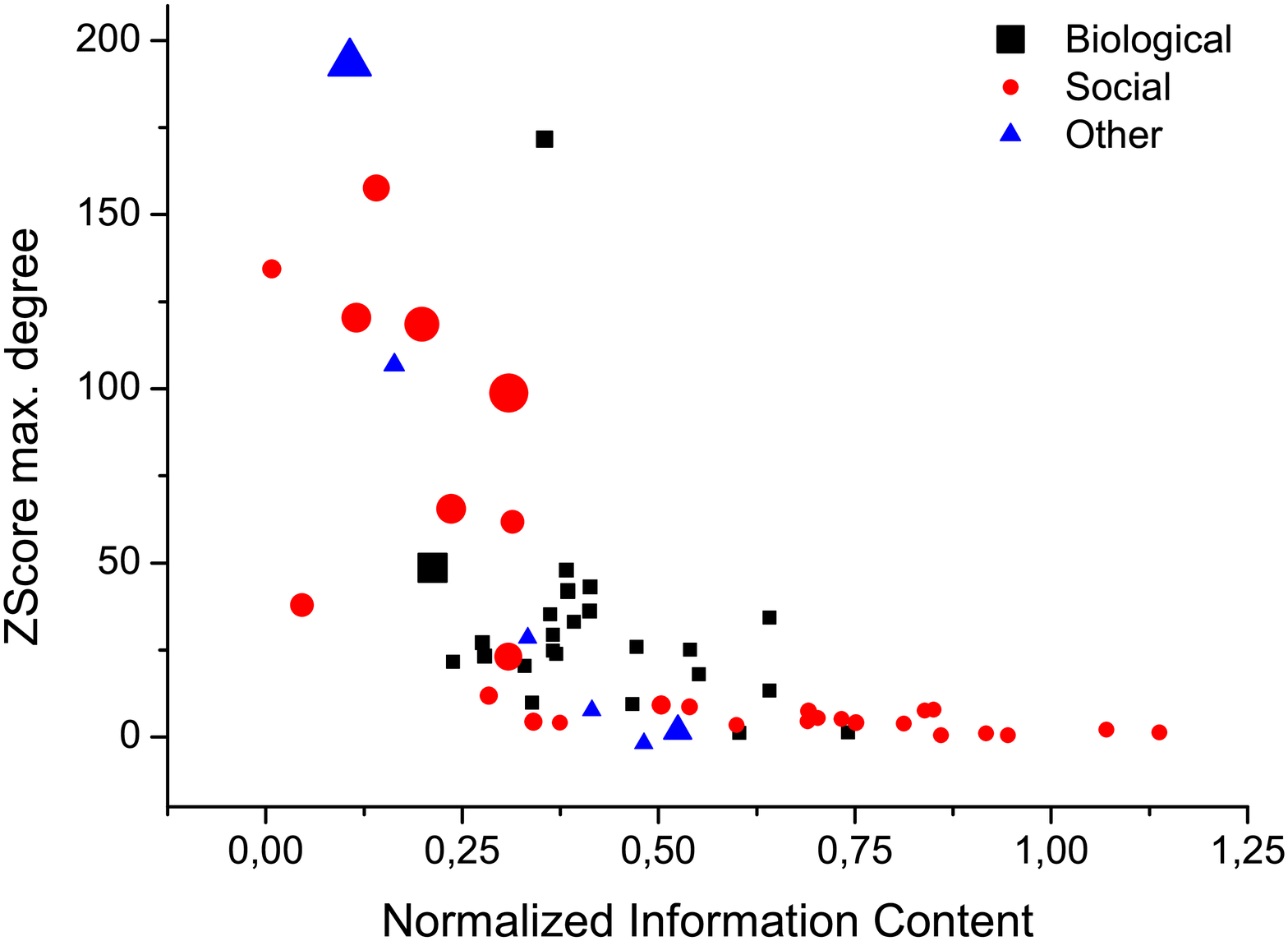}
\hspace{1cm}
\includegraphics[width=0.4\textwidth]{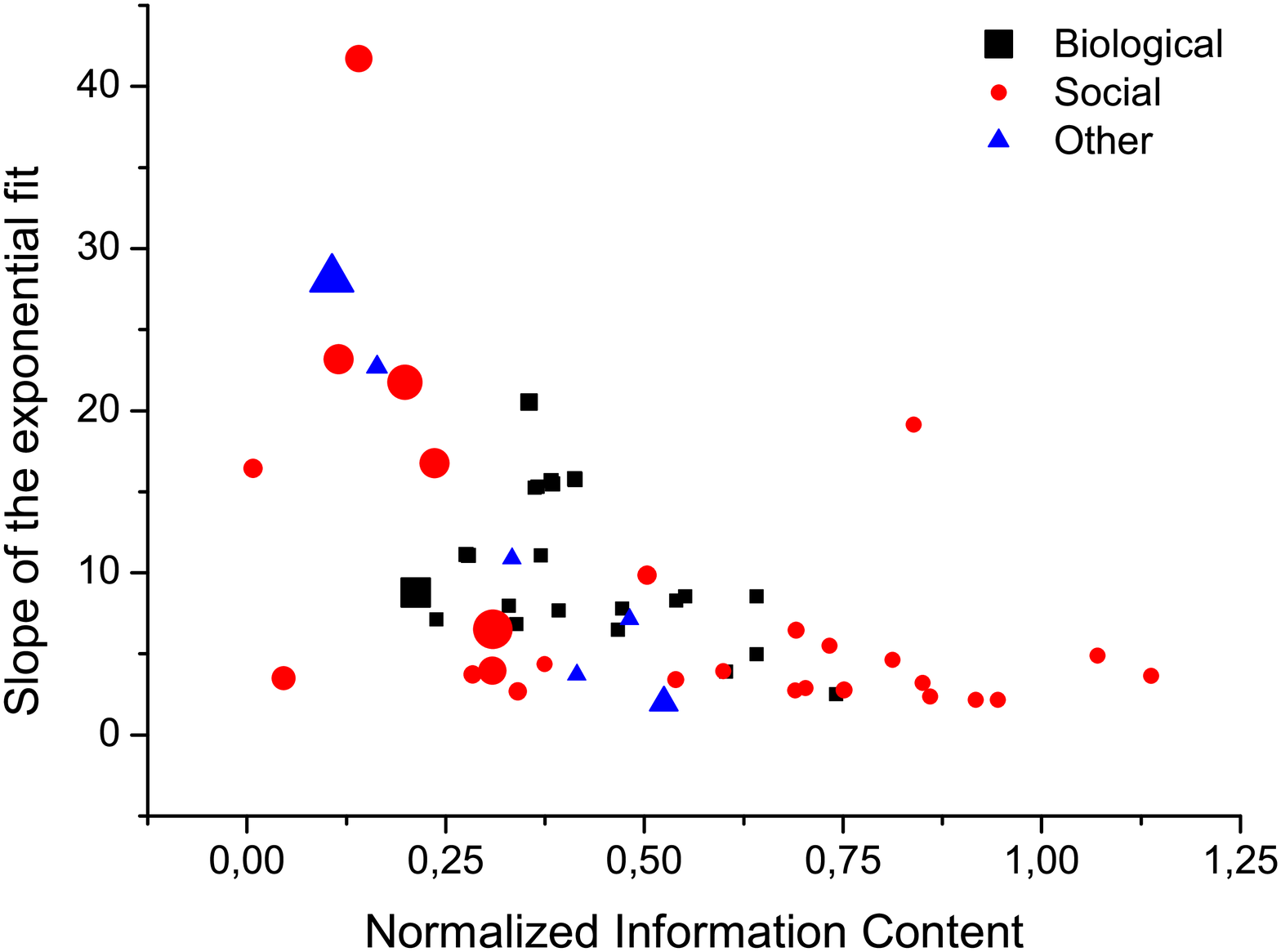}
}
\vspace{1cm}
\centering{
\includegraphics[width=0.4\textwidth]{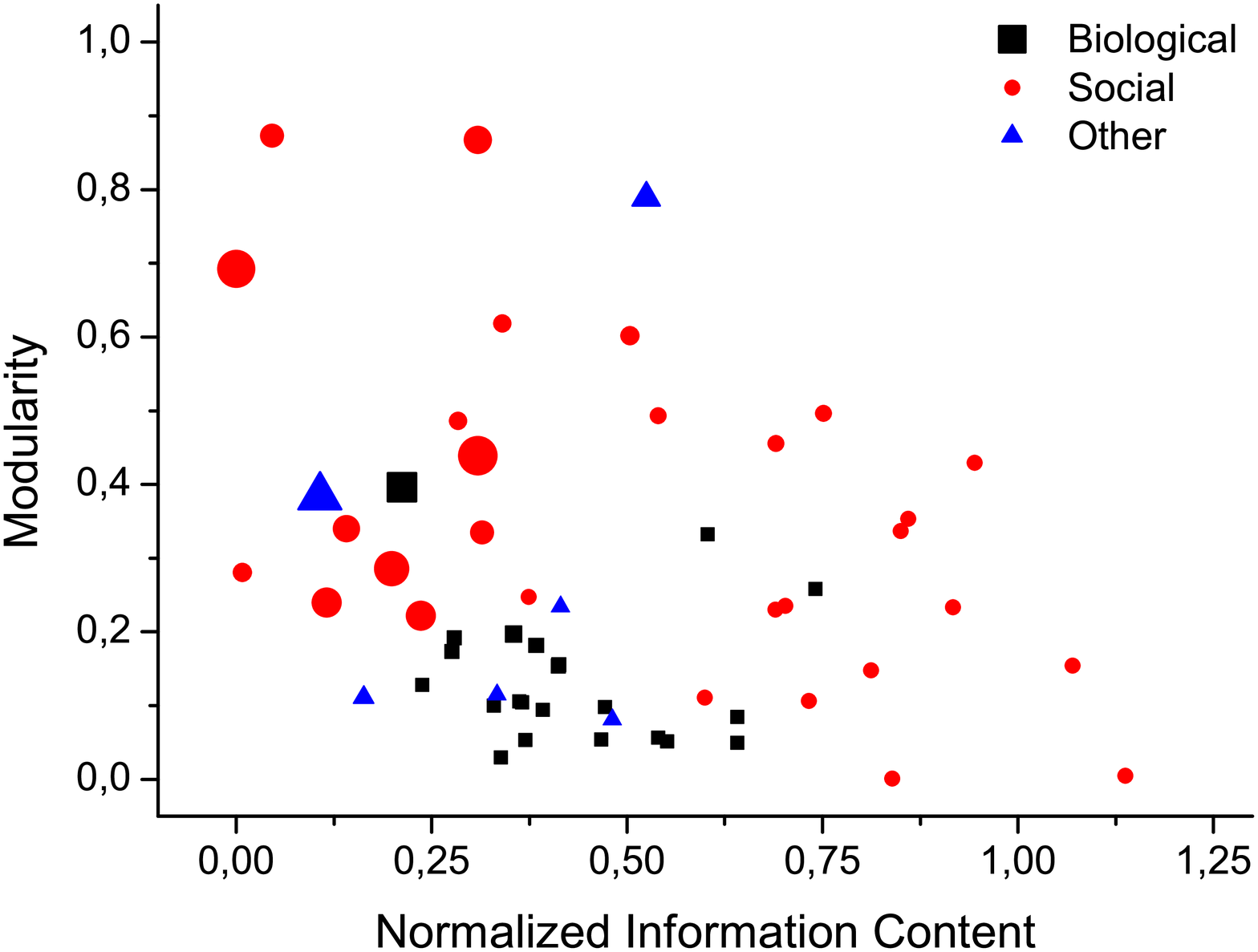}
\hspace{1cm}
\includegraphics[width=0.4\textwidth]{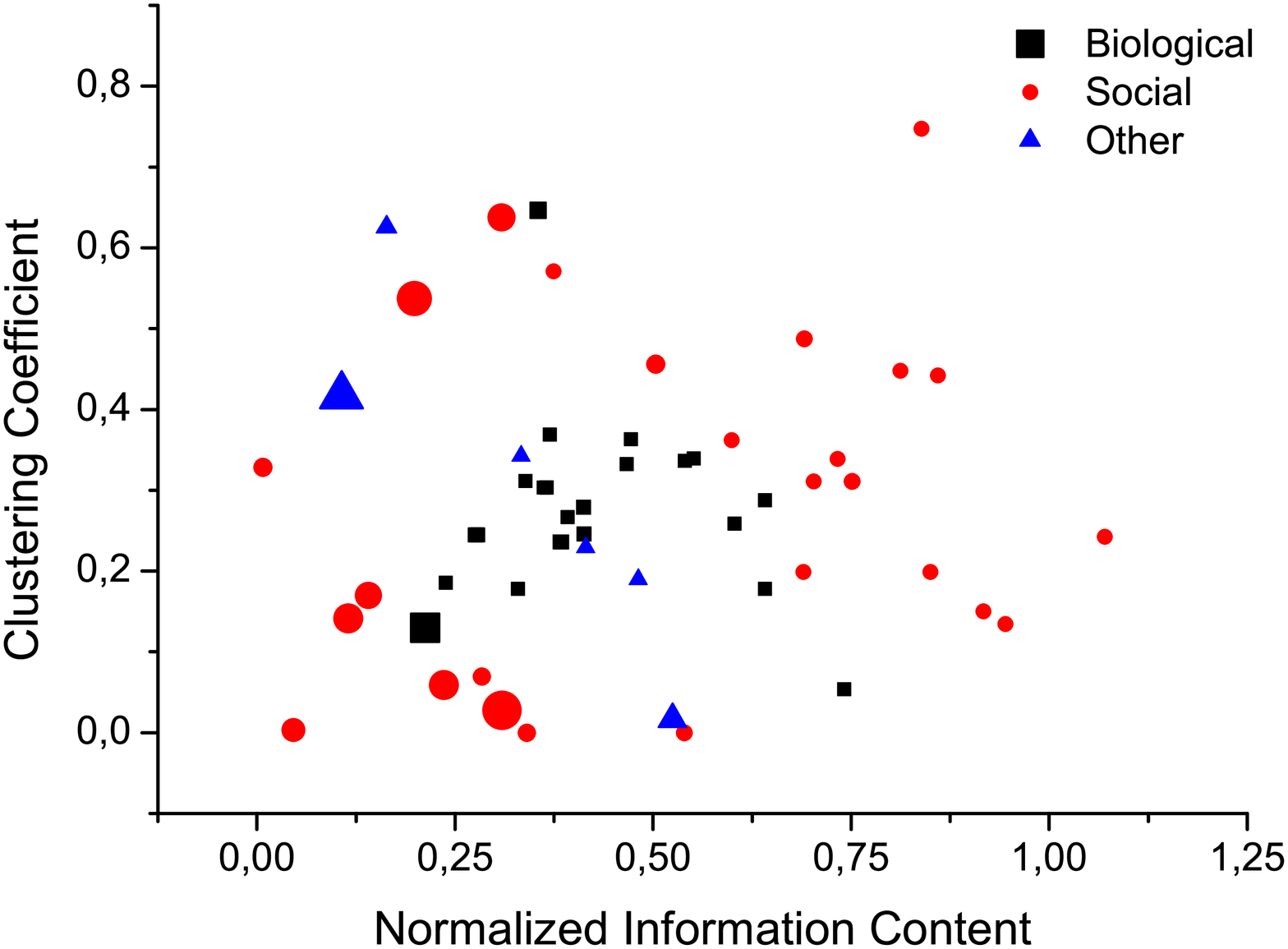}
}
\caption{(Color online) {\it Phenospaces of $55$ real networks}.
In the four panels, each network is represented by a point, whose coordinates are given by $IC_{norm}$ and a second topological metric ({\it i.e.} from left to right, top to bottom, ZScore of the maximum node degree, slope of the exponential fit of the degree distribution, modularity and clustering coefficient). Colors encode the type of system represented by each network: black squares for biological systems, red circles for social, and blue triangles for other types of systems (mainly technological); the size of each point represents the size of the corresponding network.
\label{fig:3}}
\end{figure*}

In summary, a low value of $IC_{norm}$ indicates the presence of some kind of meso-scale regularity, although it gives no information about the specific type of structure detected; in other words, one knows that a structure is present, but not if it is a modular structure, a bipartite one, {\it etc}. Thus it is natural to complement the information yielded by $IC$ with other common topological metrics. 
In order to stress this point, Fig. \ref{fig:3} presents four different phenospace of $55$ real networks, covering social, biological and technological systems \cite{datasetsA,datasetsB,datasetsC,datasetsD,datasetsE,datasetsF,datasetsG,datasetsH}. 
Each network is represented as a point in the plane, whose coordinates are given by the $IC_{norm}$ and by the value of a second topological metric, drawn from the following: ZScore of the maximum node degree, slope of the exponential fit of the degree distribution, modularity  (as calculated with the Blondel's community detection algorithm \cite{blondel2008}) and clustering coefficient.
If the pair of topological metrics considered in each phenospace were equivalent, one should expect all points to lie on a line. On the contrary, the four panels of Fig. \ref{fig:3} display a large variety of relationships. First, an inverse relationship between $IC_{norm}$ on the one hand, and ZScore of the maximum node degree (top left panel) and the slope of the exponential fit (top right panel) on the other, can be appreciated; second, modularity and clustering coefficient yield graphs in which points cover the whole plane, indicating that the information they provide is not redundant. Thus, a low {\it Information Content} cannot immediately be associated to a given meso-scale feature, but it should be complemented with different phenospace analyses.
It is also worth noticing the different behaviors corresponding to the different types of networks: social networks (red circles) cover the whole parameter space, while biological networks (black squares) seem to be bounded inside specific regions.

\begin{figure*}[!b]
\includegraphics[width=0.4\textwidth]{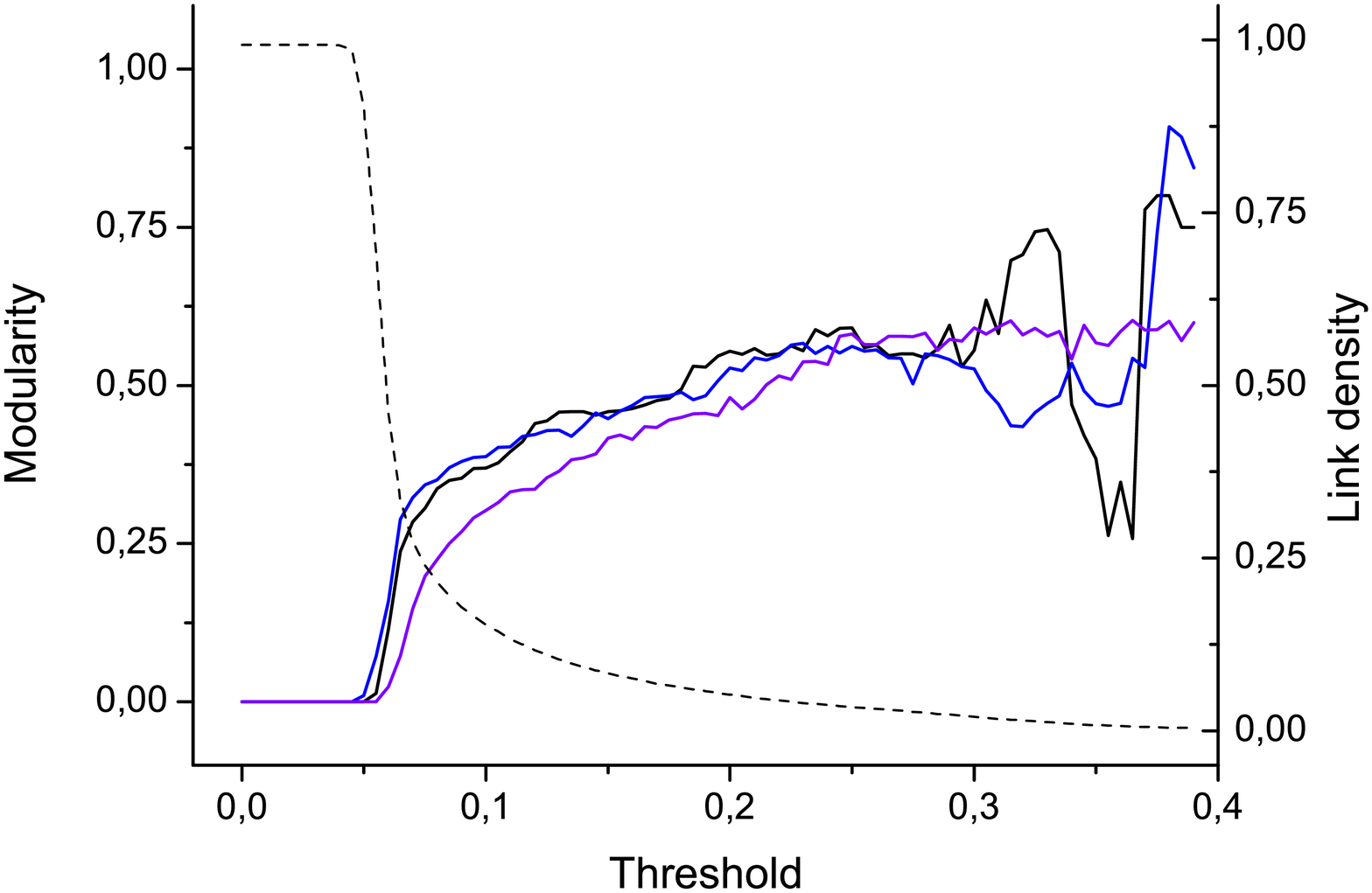}
\hspace{1cm}
\includegraphics[width=0.4\textwidth]{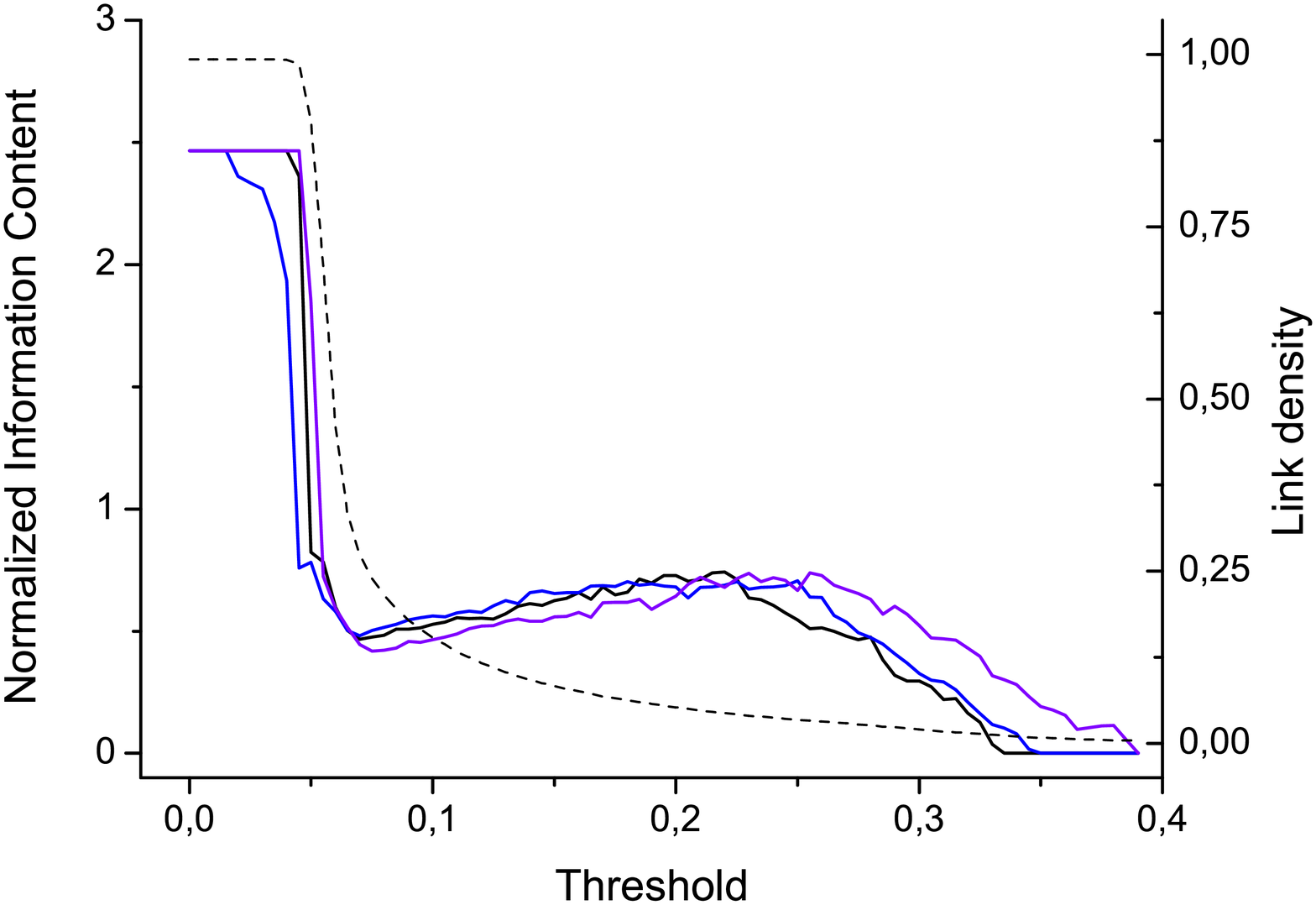}
\caption{(Color online) {\it Modularity and $IC_{norm}$ in weighted functional brain networks}.
Evolution of the modularity (Left) and of the normalized {\it Information Content} (Right) for three human brain functional networks, as a function of the applied threshold. Dotted gray lines represent the corresponding link density (right axes).
\label{fig:4}}
\end{figure*}

%

{\it Information Content} can also be used to assess the presence of different structures in weighted networks, by applying different thresholds and track how the $IC_{norm}$ evolves. As a test case, here we consider three brain functional networks \cite{bullmore2009}, obtained through magneto-encephalographic (MEG) recordings  of three healthy subjects performing a Sternberg's letter-probe task. For each subject, a weighted clique of size $148 \times 148$ was computed using the MEG time series, where the weights are given by the correlation between each pair of sensors as calculated by means of a Synchronization Likelihood (SL) algorithm \cite{stam2002}.

Fig. \ref{fig:4} reports the evolution of the modularity and of the normalized {\it Information Content} for the three subjects as a function of the applied threshold. While the former has a monotonous behavior (except for high thresholds, where the reduced amount of links results in strong fluctuations), the $IC_{norm}$ presents a clear maximum corresponding to a threshold of $0.2 - 0.25$. This region of reduced topological regularity points to a change in the structure of the networks, which is consistent with the varying fractal topology of the human brain at different synchronization thresholds \cite{bassett2006A,bassett2006B}.

%

\section{Conclusions}

In conclusion, this Letter reports on the definition of a new metric designed to assess the presence of regular meso-scale structures in complex networks - a MATLAB \copyright~implementation of the $IC$ algorithm can be found in \cite{website}. While other metrics, {\it e.g.} modularity, are defined {\it a posteriori}, that is the community structure should be detected before the calculation of the modularity of a network, {\it Information Content} can be obtained directly from the adjacency matrix. Furthermore, it is an exact metric, not requiring any optimization process whose result depends on the specific algorithm used. Finally, it enables the simultaneous assessment of different meso-scale structures, providing information complementary to standard measures. For all this, {\it Information Content} is expected to provide important benefits in tasks requiring the systematic and automatized analysis of large sets of networks, as in the case of classification tasks, for instance when a network representation is used to assess the health status of different patients \cite{zanin2012A,zanin2012B,zanin2012C}.

\end{document}